\documentclass[12pt,]{article}
\usepackage{authblk}
\usepackage{fullpage}
\usepackage{amssymb,amsmath}
\usepackage[utf8x]{inputenc}
\usepackage[T1]{fontenc}
\usepackage{siunitx}
\usepackage[version=3]{mhchem}

\usepackage[sort&compress,square,comma,numbers]{natbib}
\usepackage{setspace}
\usepackage{graphicx}
\graphicspath{{figures/}}
\makeatletter
\def\ScaleWidthIfNeeded{%
 \ifdim\Gin@nat@width>\linewidth
    \linewidth
  \else
    \Gin@nat@width
  \fi
}
\def\ScaleHeightIfNeeded{%
  \ifdim\Gin@nat@height>0.9\textheight
    0.9\textheight
  \else
    \Gin@nat@width
  \fi
}
\makeatother
\setkeys{Gin}{width=\ScaleWidthIfNeeded,height=\ScaleHeightIfNeeded,keepaspectratio}%

\DeclareMathOperator\erf{erf}

\title{Non-imaging metasurface design for collimated beam shaping}
\author[1,$\dagger$]{Kirstine E. S. Nielsen}
\author[1]{Mads A. Carlsen}
\author[1]{Xavier Zambrana-Puyalto}
\author[1,*]{S\o ren Raza}
\affil[1]{Department of Physics, Technical University of Denmark, Fysikvej, DK-2800 Kongens Lyngby, Denmark}
\affil[$\dagger$]{kiesan@dtu.dk}
\affil[*]{sraz@dtu.dk}

\date{}

\begin{document}

\maketitle

\newpage
\section*{Abstract}
Non-imaging optical lenses can shape the light intensity from incoherent sources to a desired target intensity profile, which is important for applications in lighting, solar light concentration, and optical beam shaping. Their surface curvatures are designed to ensure optimal transfer of energy from the light source to the target. The performance of such lenses is directly linked to their asymmetric freeform surface curvature, which is challenging to manufacture. Metasurfaces can mimic any surface curvature without additional fabrication difficulty by imparting a spatially-dependent phase delay using optical antennas. As a result, metasurfaces are uniquely suited to realize non-imaging optics, but non-imaging design principles have not yet been established for metasurfaces.
Here, we take an important step in connecting non-imaging optics and metasurface optics, by presenting a phase-design method for beam shaping based on the concept of optimal transport. We establish a theoretical framework that enables a collimated beam to be redistributed by a metasurface to a desired output intensity profile. The optimal transport formulation leads to metasurface phase profiles that transmit all energy from the incident beam to the output beam, resulting in an efficient beam shaping process. Through a variety of examples, we show that our approach accommodates a diverse range of different input and output intensity profiles. Last but not least, a full field simulation of a metasurface has been done to verify our phase-design framework. 
\newpage

\section*{Introduction}
Shaping light from incoherent sources, such as light-emitting diodes, has important applications in solar light concentration and indoor and outdoor illumination design. This is traditionally achieved by freeform refractive lenses using design principles from non-imaging optics~\cite{Wu2018}. Non-imaging optics focuses on the optimal transfer of energy from a light source onto a target without requiring the formation of an image of the source. The design of non-imaging freeform lenses is rooted in geometrical optics and is formulated as an inverse problem with the aim to redistribute a light source with a known intensity profile into a desired output intensity. Strategies to tackle this inverse problem can be broadly classified into two approaches~\cite{Brix2015}: iterative schemes~\cite{Ding:08,Sun:09,Fournier:10,Chen:12} and a nonlinear differential equation approach based on an optimal transport formulation~\cite{Ries:02,Wu:13b,Gutierrez:18}. The iterative schemes are based on an initial lens design that is iterated to achieve the desired beam shaping. This requires a priori knowledge of the lens shape, limiting the complexity of the beam shaping. On the other hand, the optimal transport method makes no a priori symmetry assumptions on the lens shape and has been demonstrated to work for a variety of incident light distributions and output illumination patterns~\cite{Wu:14}. This enables complex illumination patterns~\cite{Brix2015} akin to those obtained using holographic methods for coherent light sources. Noticeably, realizing such beam shaping requires manufacturing of freeform lenses with surface accuracy and smoothness close to the nanoscale to retain their designed functionality~\cite{Wang2009}. Lens manufacturing based on single point diamond turning and injection molding struggles to meet these requirements~\cite{Fang2013}. In contrast, metasurfaces offer the ability to impart any phase profile with no additional fabrication difficulty~\cite{Zhan2017,Nikolov2021} and large-scale fabrication has been demonstrated using laser printing~\cite{Zhu2018} and nanoimprint lithography~\cite{Einck2021}. 

Metasurfaces are nanostructured optical surfaces that are capable of manipulating the properties of light in a judicious manner, which has resulted in a wide range of flat optical devices~\cite{Dominguez:14,Zhou:17,Groep:20,Lawrence:20,Cai:21,Zhou:18,Weiss:22,Kim:21,Engelberg:20,Wang:17}. Their functionality is based on imparting a spatially-dependent phase delay using optical nanoantennas~\cite{Kamali2018,Chen2020} to shape the transmitted~\cite{Yu2011,Decker:15,Arbabi2015,Lalanne2017} or reflected wavefront~\cite{Pors2013}. The metasurface design process consists of specifying the phase profile needed to achieve a desired beam shaping functionality. The phase profile can be obtained either by analytical expressions~\cite{Aieta:12b}, numerical calculations using commercial ray-tracing software~\cite{Arbabi2016}, as well as holographic methods~\cite{Desiatov2015}. The latter enables almost arbitrary beam shaping~\cite{Scheuer:17}, but requires an incident light source that is coherent. While progress has been made recently~\cite{Khaidarov2020,Mukherjee2023}, shaping of incoherent light using metasurfaces has remained relatively unexplored~\cite{So2023}. Applying the concepts from non-imaging optics to the phase design of metasurfaces is a promising approach to shape incoherent light using flat optics. 

Here, we demonstrate a metasurface formulation of the optimal transport approach to determine the phase profile for one-dimensional intensity shaping of a collimated beam. The optimal transport concept ensures that all energy from the incident beam is transmitted to the output beam, resulting in a metasurface phase design with efficient beam shaping. Using the generalized law of refraction, we derive a nonlinear differential equation for the metasurface phase profile, which depends on the intensity profiles of the incident and output beams, as well as the separation distance between the metasurface and target plane. We demonstrate three beam-shaping cases with analytical solutions, and two cases which are solved numerically. The results from these five different combinations of incident and output illuminations serve to demonstrate the versatility of the method in the design of non-imaging metasurfaces. Finally, we verify our theoretical framework with full-field electromagnetic simulation. We simulate a non-imaging metasurface whose phase is the result of our model, and we compare it to its theoretical output. We quantify the differences between the theoretical solution and the full-field simulation, and we find them to be in good agreement.

\section*{Theoretical framework}
\begin{figure}[!bp]
\centering\includegraphics[width=13cm]{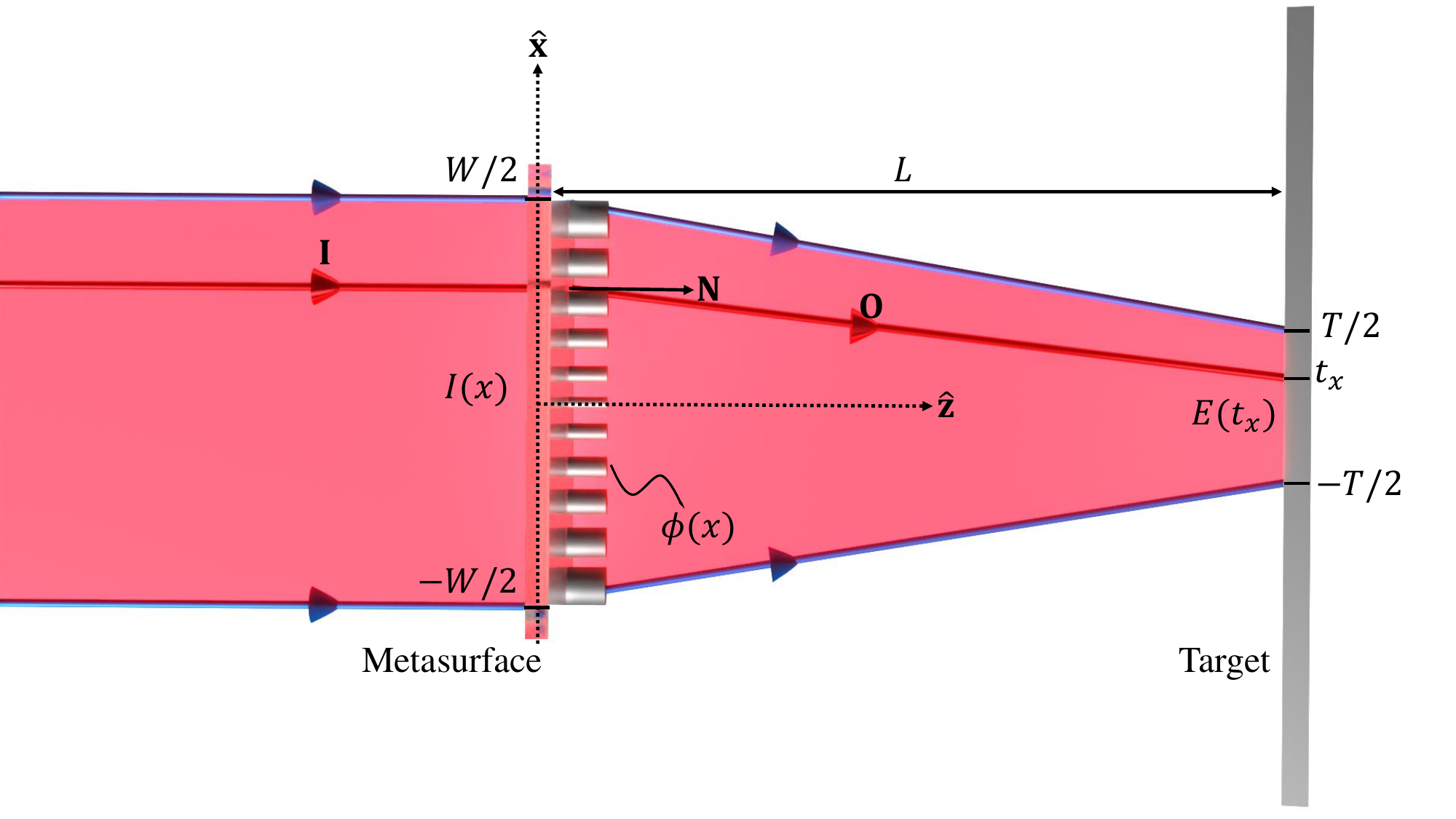}
\caption{Schematic representation of the functionality of the non-imaging metasurface. A normally-incident collimated beam $\boldsymbol{\mathrm{I}}$ with intensity profile $I(x)$ is incident on a metasurface with phase profile $\phi(x)$. The metasurface, positioned at $z = 0$, has a width of $W$ and refracts the beam onto the target plane with width $T$ and positioned a distance $L$ from the metasurface. The refracted ray $\boldsymbol{\mathrm{O}}$ hits the target plane at a specific coordinate $t_x$ to produce the desired intensity profile $E(t_x)$. The boundary rays of the beam (blue) map from the edges of the metasurface to the edges of the target plane in accordance with the edge-ray principle.}
\label{fig:CoverImage}
\end{figure}
We consider a normally-incident collimated beam of light with a one-dimensional intensity profile $I(x)$, which is redistributed by a refracting metasurface with phase profile $\phi(x)$ to a desired target illumination with intensity profile $E(t_x)$~(Fig.~\ref{fig:CoverImage}). The coordinate $t_x$ at the target is located a distance $L$ from the metasurface and is connected to the $x$-coordinate at the metasurface through the refraction caused by the metasurface phase gradient. Our aim is to determine the metasurface phase profile that maps the incident intensity profile to the desired target intensity profile. This constitutes an inverse problem, which we formulate by drawing inspiration from the concept of optimal transport in non-imaging freeform optics~\cite{Wu:13b}. 

We assume that the metasurface is a refractor, which transmits all incident light and is lossless. Metasurfaces with near unity transmission and $2\pi$ phase coverage can be realized with high-refractive-index dielectrics using Huygens'-type~\cite{Decker:15,Raza2020} and nanopost metaatoms~\cite{Arbabi2015}. Energy conservation then dictates that the power incident on the metasurface is equal to the power received at the target~\cite{Wu:13a}
\begin{align} \label{eq:EconvGlobal}
    \int_{W}I(x) \textrm{d}x = \int_{T}E(t_x)\textrm{d}t_x = \int_{W}E(t_x)\left|\frac{\partial t_x}{\partial x}\right|\textrm{d}x,
\end{align}
where $W$ and $T$ denote the widths of the metasurface and target plane, respectively, and $\left|\frac{\partial t_x}{\partial x}\right|$ is the Jacobian describing the change of variable from the target coordinate $t_x$ to the metasurface coordinate $x$. Physically, the Jacobian describes the ray expansion or contraction due to the refraction by the metasurface. Equation~(\ref{eq:EconvGlobal}) requires that the power is locally conserved and results in the differential equation
\begin{align} \label{eq:EconvLocal}
    \frac{\partial t_x}{\partial x} = \pm \frac{I(x)}{E(t_x)},
\end{align}
where we have lifted the absolute value of the Jacobian. The proper choice of sign in Eq.~(\ref{eq:EconvLocal}) is connected to the boundary conditions, which we discuss in detail later. 

Our aim now is to establish a relationship between the target coordinate $t_x$ and the metasurface phase profile $\phi(x)$. From geometrical considerations (Fig.~\ref{fig:CoverImage}), we obtain
\begin{equation}\label{eq:CoordinatesDone}
    t_x = x + L\frac{O_x}{O_z},
\end{equation}
where $\boldsymbol{\mathrm{O}}=(O_x,O_z)$ is the unit vector describing the direction of the refracted ray.
The direction of the refracted ray is determined by its wave vector $\boldsymbol{\mathrm{k_r}} = (k_x,k_z)$, which is found from the generalized law of refraction. For a normally-incident beam, the generalized law of refraction in one dimension can be expressed as $\frac{\partial\phi}{\partial x} = k_x$, where $\phi(x)$ is the phase imparted by the metasurface~\cite{Aieta:12a, Moreno2020, Castaneda-Almanza:22}. The direction of the refracted ray is therefore determined by the metasurface phase gradient as
\begin{equation} \label{eq:Ovec}
    \boldsymbol{\mathrm{O}} = \frac{\boldsymbol{\mathrm{k_r}}}{|\boldsymbol{\mathrm{k_r}}|} = \frac{1}{k}(k_x,k_z) = \left(\frac{\partial\Phi}{\partial x},\sqrt{1-\left(\frac{\partial\Phi}{\partial x}\right)^2}\right).
\end{equation}
Here, we have introduced a scaled wavelength-independent phase $\Phi(x) =\phi(x)/(n_0k_0)$, where $n_0$ is the refractive index of the medium between the metasurface and the target plane, and $k_0=2\pi/\lambda$ is the vacuum wavenumber. Combining Eqs.~(\ref{eq:CoordinatesDone}-\ref{eq:Ovec}), we arrive at an expression for the phase gradient in terms of the target coordinate
\begin{align}\label{eq:gradient}
    \frac{\partial\Phi}{\partial x} = \frac{t_x-x}{\sqrt{L^2+(t_x-x)^2}}.
\end{align}
Equations~(\ref{eq:EconvLocal}) and (\ref{eq:gradient}) constitute the basis of our theoretical framework. The solution to Eq.~(\ref{eq:EconvLocal}) provides the target coordinate $t_x(x)$ for a desired beam shaping functionality, which is specified through the incident intensity $I(x)$ and output intensity $E(t_x)$. Once the target coordinate is obtained, Eq.~(\ref{eq:gradient}) provides the phase gradient, from which we obtain the metasurface phase profile through integration. 

\subsection*{Boundary conditions}
\begin{figure}[!tbp]
\centering\includegraphics[width=13cm]{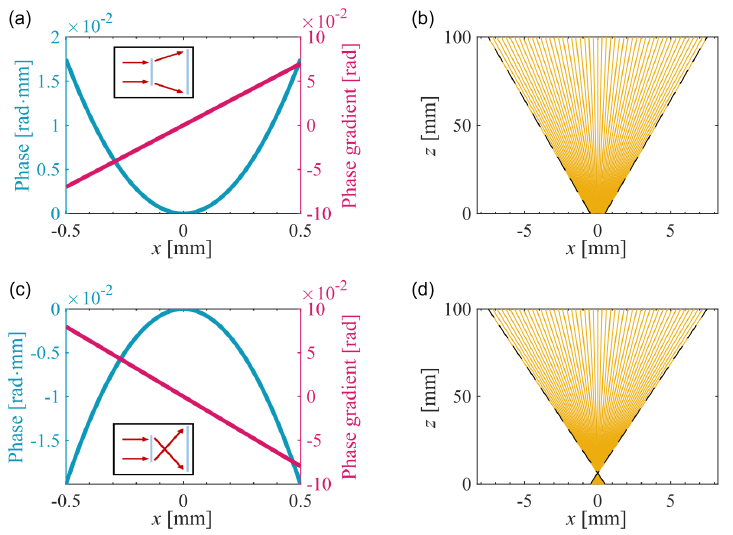}
\caption{(a,c) Phase profiles and phase gradients, and (b,d) resulting ray plots for a diverging non-imaging metasurface. In (c-d) the boundary conditions are reversed relative to (a-b), as illustrated in the insets. In (b,d) the dashed black lines indicate the boundary rays. The metasurface width, target plane width and metasurface-to-target separation are $W=1$~mm, $T=15$~mm, and $L=10$~cm, respectively.}
\label{fig:Defocus_BC}
\end{figure}

The proper choice of sign in the differential equation in Eq.~(\ref{eq:EconvLocal}) is determined by the boundary conditions, which set the trajectories of the light rays impinging on the edges of the metasurface. In non-imaging optics, these boundary conditions are given by the edge-ray principle. The edge-ray principle states that a mapping of all rays from the edge of the source distribution should be directed to the edge of the target illumination distribution, in order to ensure that all rays fall within the target~\cite{Ries:94}. In our case, this yields two possible sets of boundary conditions, as illustrated in the insets of Fig.~\ref{fig:Defocus_BC}(a,c),
\begin{align}\label{Eq:BCanalytical}
   \text{1st set:\,\,\,\,} t_x\left(x=\pm \frac{W}{2}\right) = \pm \frac T 2, \\\label{Eq:BCreverse}
   \text{2nd set:\,\,\,\,}t_x\left(x=\pm \frac{W}{2}\right) = \mp \frac T 2.
\end{align}

We illustrate the relation between the choice of boundary conditions and the sign in Eq.~(\ref{eq:EconvLocal}) through the beam shaping of a diverging lens~(Fig.~\ref{fig:Defocus_BC}). In our framework, this is described by a target width larger than the metasurface width ($T>W$) as well as incident and output beams of constant intensities, i.e., $I(x)=I_0$ and $E(t_x)=E_0$. From global energy conservation [Eq.~(\ref{eq:EconvGlobal})] we obtain $E_0 = \frac{W}{T}I_0$, and using this to solve Eq.~(\ref{eq:EconvLocal}) yields
\begin{align}\label{eq:tx_contant_to_constant}
    t_x(x) = \pm \frac{T}{W}x
\end{align}

Here, we observe that the positive ($+$) sign can only satisfy the first set of boundary conditions [Eq.~(\ref{Eq:BCanalytical})], while the negative ($-$) only satisfies the second set [Eq.~(\ref{Eq:BCreverse})]. Consequently, the choice of sign in the differential equation [Eq.~(\ref{eq:EconvLocal})] comes with a unique set of boundary conditions. The two different approaches achieve the same intensity profile at the target plane, however, the metasurface phase profiles differ substantially. By using the relation in Eq.~(\ref{eq:gradient}) and setting $\Phi(x=0) = 0$, we find the following analytical expressions for the two phase profiles

\begin{equation}
    \Phi_{\pm}(x) = \pm \left( \frac{T}{W} \mp 1 \right)^{-1} \left[ \sqrt{ L^2 + x^2 \left( \frac{T}{W} \mp 1 \right )^2 } \mp L \right]. \\
\end{equation}

In the case of a positive sign, the phase profile of the metasurface corresponds to a regular diverging lens, which spreads out the incident beam evenly to achieve a constant intensity at the target plane [Fig.~\ref{fig:Defocus_BC}(a,b)]. In contrast, the choice of a negative sign leads to a focusing phase profile with a focal length $f$ smaller than the distance to the target plane ($f<L$) [Fig.~\ref{fig:Defocus_BC}(c,d)]. After the initial focusing at the focal point, the rays spread out to result in the same constant intensity at the target plane as for the diverging lens. It is therefore of value to note that even though the phase profiles and ray trajectories differ, both approaches are equally valid within our theoretical framework. Although both approaches result in the same beam shaping, choosing the negative sign may lead to a lens design with a high numerical aperture, which may be difficult to realize. Thus we choose to focus on the approach described by the positive sign in Eq.~(\ref{eq:EconvLocal}) along with the first set of boundary conditions [Eq.~(\ref{Eq:BCanalytical})], i.e.,
\begin{align}\label{eq:FinalModel}
    \frac{\partial t_x}{\partial x} = \frac{I(x)}{E(t_x)},\,\,\, \text{with BC:\,\,\,\,} t_x\left(x=\pm \frac{W}{2}\right) = \pm \frac T 2
\end{align}

\begin{figure}[!tbp]
\centering\includegraphics[width=13cm]{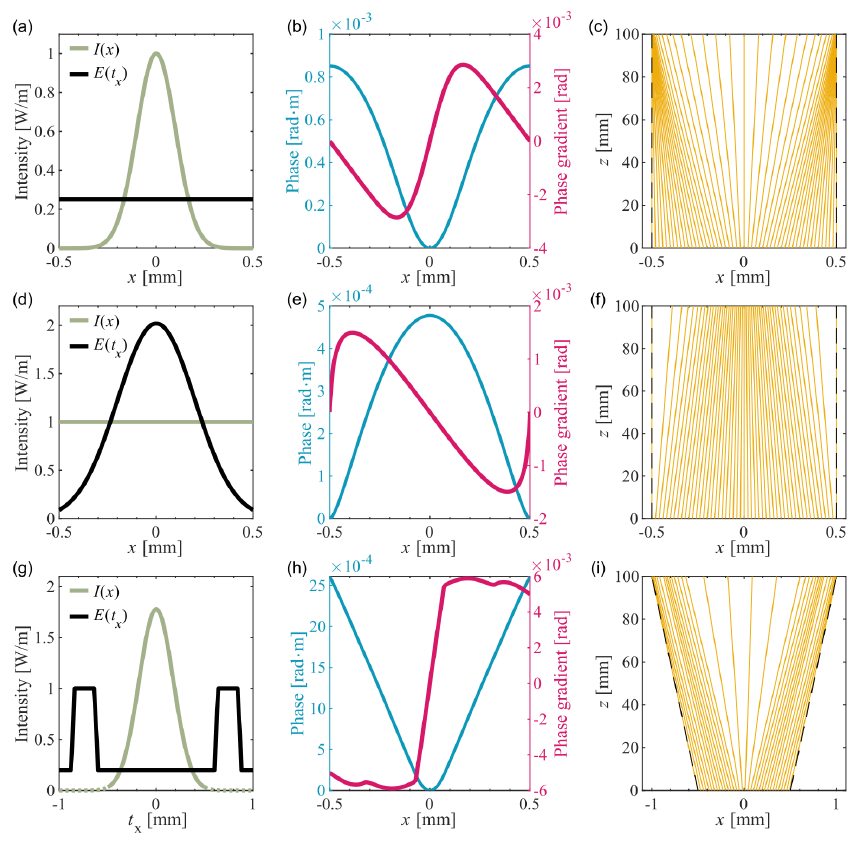}
\caption{Three examples of beam shaping: (a-c) Gaussian-to-constant mapping, (d-f)~constant-to-Gaussian mapping, and (g-i)~Gaussian-to-square mapping. The Gaussian-to-constant mapping (a) exhibits a concave phase profile (b) that spreads the beam (c). The constant-to-Gaussian mapping (d) displays a convex phase profile (e) consistent with a focusing effect (f). In both cases the metasurface and target plane have the same width ($T = W$). The Gaussian-to-square mapping (g) displays overall defocusing, since the target plane is wider than the metasurface ($T=2W$), as well as local focusing to generate the square target profiles (h,i). For all cases, the metasurface width and metasurface-to-target separation are $W=1$~mm and $L=100$~mm, respectively.}
\label{fig:3examples}
\end{figure}
\section*{Beam shaping}
Having established the theoretical framework, we now determine the metasurface phase profiles for three different beam shaping examples. An advantage of our theoretical framework is that significant analytical progress can be made for certain beam shaping cases. In particular, shaping a Gaussian input intensity profile to a constant output intensity (and vice versa) can be solved analytically. For the Gaussian-to-constant mapping, we use the input intensity profile $I(x) = I_0\exp \left({-2x^2/w_0^2}\right)$ with $w_0$ being the waist radius~\cite{FundamentalsOfPhotonics}, while the target intensity profile is constant $E(t_x) = E_0$, see Fig.~\ref{fig:3examples}a. In this case we find that the target coordinate is given by
\begin{align}\label{eq:tx_gaussian_to_constant}
    t_x(x) = \frac{T}{2}\frac{\erf\left(\frac{\sqrt{2}x}{w_0}\right)}{\erf\left(\frac{\sqrt{2}W}{2w_0}\right)},
\end{align}
where $\erf$ denotes the error function. We visualize the Gaussian-to-constant shaping in Fig.~\ref{fig:3examples}(b,c) with the width of the Gaussian set to $w_0=0.2$~mm and the metasurface and target widths being equal ($T=W=1$~mm). We find a concave-type phase profile (Fig.~\ref{fig:3examples}b), which is reasonable as the beam shaping taking place is similar to a diverging lens (see Fig.~\ref{fig:Defocus_BC}a). The rays are directed towards the perimeter of the target plane, resulting in a flattening of the Gaussian beam shape (Fig.~\ref{fig:3examples}c). It is however worth noting that the phase gradient differs from a purely diverging lens as the edge rays must be mapped to the edge of the target. Since the target and metasurface widths are equal, this forces the phase gradient to go to zero at the edges of the metasurface (see Eq.~(\ref{eq:gradient})).

For a constant-to-Gaussian mapping, the input intensity profile is constant $I(x) = I_0$, while the target intensity has a Gaussian profile $E(t_x) = E_0 \exp\left(-2t_x^2/w_0^2\right)$ with $w_0 = 0.4$~mm, see Fig.~\ref{fig:3examples}d. Here, we find that the target coordinate is 
\begin{align}
    t_x(x) = \frac{w_0}{\sqrt{2}} \erf^{-1} \left(\frac{2}{W}\text{erf}\left(\frac{\sqrt{2}T}{2w_0}\right)x\right),
\end{align}
where $\erf^{-1}$ is the inverse error function. The phase profile has a convex shape (Fig.~\ref{fig:3examples}e), which leads to focusing of the incident beam to the desired Gaussian target (Fig.~\ref{fig:3examples}f). 

As a final demonstration, we consider a Gaussian incident beam that is mapped to a target profile consisting of two square bumps with a ratio of 5:1 between the maximum and minimum intensity~(Fig.~\ref{fig:3examples}g). The minimum output intensity is set to a low but finite value, as zero-valued output intensities violates the local energy conservation dictated by Eq.~(\ref{eq:EconvLocal}). The width of the target plane is set to twice the width of the metasurface ($T=2W=2$~mm). To determine the phase profile, we solve the theoretical model numerically. To this end, it is worth noting that the first-order nature of the differential equation in Eq.~(\ref{eq:FinalModel}) requires in principle only a single boundary condition to determine a solution. To satisfy both boundary conditions dictated by the edge-ray principle [Eq.~(\ref{eq:FinalModel})], we must limit the solution space to beam shaping of symmetric input and output intensity profiles. This is a consequence of the one-dimensional nature of our theoretical framework. This constraint can be released by extending the framework to two dimensions, since boundary rays can then map to arbitrary points on the edge of the target profile and thereby enable asymmetric intensity profiles, as demonstrated with freeform optics~\cite{Wu:13b}. Nonetheless, taking into account the symmetry condition of our framework comes with the advantage that Eq.~(\ref{eq:FinalModel}) can be solved as an initial value problem. We therefore implemented a numerical solver using a built-in initial value solver in MATLAB, which is based on the Runge--Kutta method. We use the left boundary as the initial condition, i.e., $t_x(-W/2) = -T/2$, and observe that the right boundary condition is automatically satisfied (Fig.~\ref{fig:3examples}i).
The resulting phase profile and phase gradient contain both diverging and focusing features (Fig.~\ref{fig:3examples}h). In particular, the phase gradient consists of an overall positive slope, which spreads the input beam to accommodate the wider target plane (as also seen in Fig.~\ref{fig:Defocus_BC}a), as well as two local negatively sloped features that serve to focus the input beam into the two square target profiles~(Fig.~\ref{fig:3examples}h). The ray tracing shows that the boundary conditions are satisfied, and that the designed beam shaping is achieved (Fig.~\ref{fig:3examples}i).

\section*{Metasurface design and verification}
\begin{figure}[!tbp]
\centering\includegraphics[width=13cm]{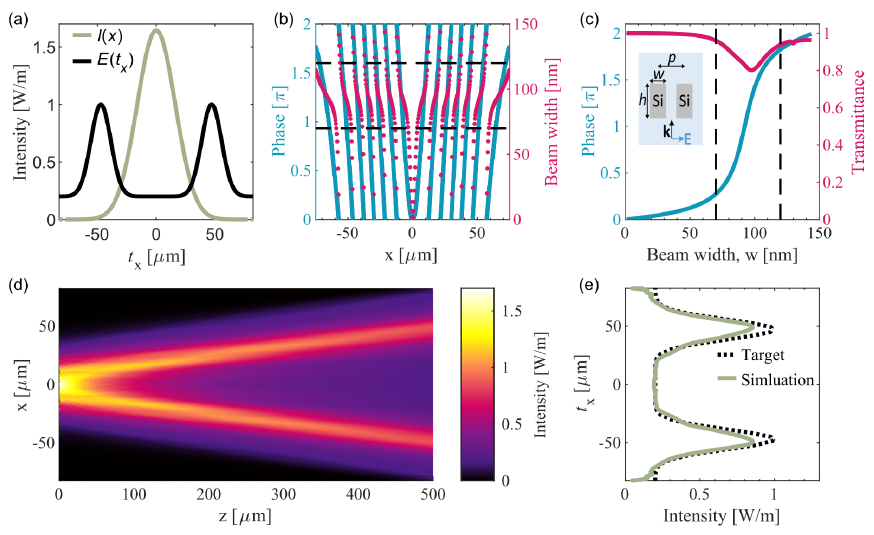}
\caption{(a) Input Gaussian intensity and target double focal spots. (b) Phase profile at the wavelength $\lambda = 660$~nm along with corresponding beam widths determined from the look-up table presented in (c). (c) Simulated look-up table for 2D periodic silicon nanobeams in air ($n_\textrm{Si}=3.84+i0.02$). The period and height are set to $p = h = 0.5\lambda$. (d) Full-field simulation result of the intensity of the output beam. The metasurface is located at $z=0$, and centered around $x=0$. (e) Simulated intensity evaluated at $z = L = 500$~\textmu m (green) and the target output intensity (dashed). The metasurface and target planes have widths of $W = 150$~\textmu m and $T = 1.1W$, respectively.}
\label{fig:Metasurface}
\end{figure}

We verify the feasibility of our theoretical model by performing a full-field 2D simulation (COMSOL Multiphysics~6.0) of a miniaturized metasurface. The metasurface is designed to split an incident Gaussian beam into two focused Gaussian profiles~(Fig.~\ref{fig:Metasurface}a). Our theoretical model and its numerical implementation determines a wavelength-independent scaled phase profile for the desired beam shaping. We retrieve the actual phase profile of the metasurface at an operating wavelength of $\lambda=660$~nm by the operation $\phi(x) = \Phi(x)n_02\pi/\lambda$ with $n_0=1$ (Fig.~\ref{fig:Metasurface}b). The phase profile is then discretized according to the look-up table based on silicon nanobeams in air (Fig.~\ref{fig:Metasurface}c).
The look-up table is generated using normally-incident plane waves and periodic boundary conditions, and applies for silicon nanobeams with a height $h = 0.5\lambda$ placed in an array with a period $p = 0.5\lambda$. The entire metasurface is constructed from 455 nanobeam elements (located at $z=0$ in Fig.~\ref{fig:Metasurface}d), resulting in a metasurface width of $150$~\textmu m. We remark that our framework is rooted in geometrical optics and does not take diffraction into account, which is appropriate for shaping of incoherent light. However, in the full-field simulation, diffraction is present due to the interaction of the finite-sized metasurface and the infinite extent of the incident Gaussian beam. To mimic incoherent light and minimize diffraction, the  waist radius of the Gaussian beam $w_0$ is carefully chosen to ensure that the intensity of the beam is near zero at the edges of the metasurface.
Simultaneously, the width of the beam also needs to be large enough to ensure a depth of focus that generates a beam profile that is collimated throughout the simulation domain (in the absence of the metasurface). From these considerations, we find that $w_0 = 32.5$~\textmu m is a suitable waist radius. This trade-off is mainly due to computational limitations on the size of the simulation domain. We believe that an experimental demonstration based on collimated light from, e.g., a light-emitting diode, combined with a significantly larger metasurface would lift this constraint.

The propagation of the output beam and a comparison between the target output intensity and simulated intensity are shown in Fig.~\ref{fig:Metasurface}d and Fig.~\ref{fig:Metasurface}e, respectively.
The transmission through the silicon nanobeams dips below unity for beams with a width in the range 70-120 nm~(Fig.\ref{fig:Metasurface}c, dashed lines). The assumption in our theoretical model, that the metasurface has unity transmission is thus not met for most of the metasurface~(Fig.\ref{fig:Metasurface}b, dashed lines). Nonetheless, the shape of the simulated and target intensities are in good agreement. We observe an intensity ratio of 4.25:1 and a peak to peak distance of $94$~\textmu m in the simulated profile, and a slight asymmetry in the peaks. The target profile has a peak separation of $98$~\textmu m and a 5:1 intensity ratio, and symmetric Gaussian peaks. Despite the previously discussed losses in the nanobeams, we observe that $91\%$  of the incident power is transmitted in the simulation. This result demonstrates the feasibility of our theoretical model for determining the phase profile for beam shaping applications, as well as how a fairly complex phase profile can be implemented with a metasurface.

\section*{Conclusion}
In summary, we have demonstrated a theoretical optimal transport framework for obtaining the phase profile of a metasurface for one-dimensional beam shaping with collimated incident light. We show how the resulting nonlinear differential equation can be solved analytically in some cases, and numerically using a Runge--Kutta method in more complex cases. Five specific cases are investigated to demonstrate the versatility of the method within the design of non-imaging metasurfaces, and the method is verified with full-field simulations.
We add that although we have only shown examples with collimated beams, our framework can be expanded to describe point-like light sources as well. Thus we believe our work is an important step towards manipulating incoherent light with metasurfaces.

\subsection*{Funding}
S.~R. acknowledges support by the Independent Research Funding Denmark (7026-00117B). 

\subsection*{Acknowledgments}
We thank Hugh Simons and Gor Nahapetyan for assistance with the numerical implementation.

\subsection*{Disclosures}
The authors declare no conflicts of interest.

\subsection*{Data availability} 
Data underlying the results presented in this paper are not publicly available at this time but may be obtained from the authors upon reasonable request.


\bibliography{References.bib}

\newcommand{\noopsort}[1]{} \newcommand{\printfirst}[2]{#1}
  \newcommand{\singleletter}[1]{#1} \newcommand{\switchargs}[2]{#2#1}
\begin{thebibliography}{10}

\bibitem{Wu2018}
R.~Wu, Z.~Feng, Z.~Zheng, R.~Liang, P.~Ben{\'{i}}tez, J.~C. Mi{\~{n}}ano, and
  F.~Duerr, ``Design of freeform illumination optics,'' {\em Laser Photonics
  Rev.}, vol.~12, no.~7, p.~1700310, 2018.

\bibitem{Brix2015}
K.~Brix, Y.~Hafizogullari, and A.~Platen, ``{Designing illumination lenses and
  mirrors by the numerical solution of Monge–Amp{\`{e}}re equations},'' {\em
  J. Opt. Soc. Am. A}, vol.~32, no.~11, pp.~2227--2236, 2015.

\bibitem{Ding:08}
Y.~Ding, X.~Liu, Z.~Zheng, and P.~Gu, ``Freeform {LED} lens for uniform
  illumination,'' {\em Opt. Express}, vol.~16, pp.~12958--12966, Aug 2008.

\bibitem{Sun:09}
L.~Sun, S.~Jin, and S.~Cen, ``Free-form microlens for illumination
  applications,'' {\em Appl. Opt.}, vol.~48, pp.~5520--5527, Oct 2009.

\bibitem{Fournier:10}
F.~R. Fournier, W.~J. Cassarly, and J.~P. Rolland, ``Fast freeform reflector
  generation using source-target maps,'' {\em Opt. Express}, vol.~18,
  pp.~5295--5304, Mar 2010.

\bibitem{Chen:12}
J.-J. Chen, T.-Y. Wang, K.-L. Huang, T.-S. Liu, M.-D. Tsai, and C.-T. Lin,
  ``Freeform lens design for {LED} collimating illumination,'' {\em Opt.
  Express}, vol.~20, pp.~10984--10995, May 2012.

\bibitem{Ries:02}
H.~Ries and J.~Muschaweck, ``Tailored freeform optical surfaces,'' {\em J. Opt.
  Soc. Am. A}, vol.~19, pp.~590--595, Mar 2002.

\bibitem{Wu:13b}
R.~Wu, L.~Xu, P.~Liu, Y.~Zhang, Z.~Zheng, H.~Li, and X.~Liu, ``Freeform
  illumination design: a nonlinear boundary problem for the elliptic
  {M}onge-{A}mpere equation,'' {\em Opt. Lett.}, vol.~38, pp.~229--231, Jan
  2013.

\bibitem{Gutierrez:18}
C.~E. Guti\'{e}rrez and L.~Pallucchini, ``Reflection and refraction problems
  for metasurfaces related to {M}onge-{A}mpere equations,'' {\em J. Opt. Soc.
  Am. A}, vol.~35, pp.~1523--1531, Sep 2018.

\bibitem{Wu:14}
R.~Wu, P.~Ben\'{i}tez, Y.~Zhang, and J.~C. {Mi\~{n}ano}, ``Influence of the
  characteristics of a light source and target on the {M}onge-{A}mpere equation
  method in freeform optics design,'' {\em Opt. Lett.}, vol.~39, pp.~634--637,
  Feb 2014.

\bibitem{Wang2009}
K.~Wang, S.~Liu, F.~Chen, Z.~Liu, and X.~Luo, ``{Effect of manufacturing
  defects on optical performance of discontinuous freeform lenses},'' {\em Opt.
  Express}, vol.~17, no.~7, pp.~5457--5465, 2009.

\bibitem{Fang2013}
F.~Fang, X.~Zhang, A.~Weckenmann, G.~Zhang, and C.~Evans, ``{Manufacturing and
  measurement of freeform optics},'' {\em CIRP Ann.}, vol.~62, no.~2,
  pp.~823--846, 2013.

\bibitem{Zhan2017}
A.~Zhan, S.~Colburn, C.~M. Dodson, and A.~Majumdar, ``Metasurface freeform
  nanophotonics,'' {\em Sci. Rep.}, vol.~7, p.~1673, May 2017.

\bibitem{Nikolov2021}
D.~K. Nikolov, A.~Bauer, F.~Cheng, H.~Kato, A.~N. Vamivakas, and J.~P. Rolland,
  ``{Metaform optics: Bridging nanophotonics and freeform optics},'' {\em Sci.
  Adv.}, vol.~7, no.~eabe5112, p.~eabe5112, 2021.

\bibitem{Zhu2018}
X.~Zhu, M.~{Keshavarz Hedayati}, S.~Raza, U.~Levy, N.~A. Mortensen, and
  A.~Kristensen, ``{Digital resonant laser printing: Bridging nanophotonic
  science and consumer products},'' {\em Nano Today}, vol.~19, pp.~7--10, 2018.

\bibitem{Einck2021}
V.~J. Einck, M.~Torfeh, A.~McClung, D.~E. Jung, M.~Mansouree, A.~Arbabi, and
  J.~J. Watkins, ``{Scalable Nanoimprint Lithography Process for Manufacturing
  Visible Metasurfaces Composed of High Aspect Ratio TiO2 Meta-Atoms},'' {\em
  ACS Photonics}, vol.~8, no.~8, pp.~2400--2409, 2021.

\bibitem{Dominguez:14}
R.~Paniagua-Domínguez, Y.~F. Yu, E.~Khaidarov, S.~Choi, V.~Leong, R.~M.
  Bakker, X.~Liang, Y.~H. Fu, V.~Valuckas, L.~A. Krivitsky, and A.~I.
  Kuznetsov, ``A metalens with a near-unity numerical aperture,'' {\em Nano
  Letters}, vol.~18, pp.~2124--2132, Mar 2014.

\bibitem{Zhou:17}
Z.~Zhou, J.~Li, R.~Su, B.~Yao, H.~Fang, K.~Li, L.~Zhou, J.~Liu, D.~Stellinga,
  C.~P. Reardon, T.~F. Krauss, and X.~Wang, ``Efficient silicon metasurfaces
  for visible light,'' {\em ACS Photonics}, vol.~4, pp.~544--551, Jan 2017.

\bibitem{Groep:20}
J.~van~de Groep, J.-H. Song, U.~Celano, Q.~Li, P.~G. Kik, and M.~L. Brongersma,
  ``Exciton resonance tuning of an atomically thin lens,'' {\em Nature
  Photonics}, vol.~14, pp.~426--430, Jul 2020.

\bibitem{Lawrence:20}
M.~Lawrence, D.~R. Barton, J.~Dixon, J.-H. Song, J.~van~de Groep, M.~L.
  Brongersma, and J.~A. Dionne, ``High quality factor phase gradient
  metasurfaces,'' {\em Nature Nanotechnology}, vol.~15, pp.~956--961, Nov 2020.

\bibitem{Cai:21}
Z.~Cai, Y.~Deng, C.~Wu, C.~Meng, Y.~Ding, S.~I. Bozhevolnyi, and F.~Ding,
  ``Dual-functional optical waveplates based on gap-surface plasmon
  metasurfaces,'' {\em Advanced Optical Materials}, vol.~9, no.~11, p.~2002253,
  2021.

\bibitem{Zhou:18}
Y.~Zhou, I.~I. Kravchenko, H.~Wang, J.~R. Nolen, G.~Gu, and J.~Valentine,
  ``Multilayer noninteracting dielectric metasurfaces for multiwavelength
  metaoptics,'' {\em Nano Letters}, vol.~18, no.~12, pp.~7529--7537, 2018.

\bibitem{Weiss:22}
A.~Weiss, C.~Frydendahl, J.~Bar-David, R.~Zektzer, E.~Edrei, J.~Engelberg,
  N.~Mazurski, B.~Desiatov, and U.~Levy, ``Tunable metasurface using thin-film
  lithium niobate in the telecom regime,'' {\em ACS Photonics}, vol.~9, no.~2,
  pp.~605--612, 2022.

\bibitem{Kim:21}
I.~Kim, R.~J. Martins, J.~Jang, T.~Badloe, S.~Khadir, H.-Y. Jung, H.~Kim,
  J.~Kim, P.~Genevet, and J.~Rho, ``Nanophotonics for light detection and
  ranging technology,'' {\em Nature Nanotechnology}, vol.~16, pp.~508--524, May
  2021.

\bibitem{Engelberg:20}
J.~Engelberg and U.~Levy, ``The advantages of metalenses over diffractive
  lenses,'' {\em Nature Communications}, vol.~11, p.~1991, Apr 2020.

\bibitem{Wang:17}
S.~Wang, P.~C. Wu, V.-C. Su, Y.-C. Lai, C.~H. Chu, J.-W. Chen, S.-H. Lu,
  J.~Chen, B.~Xu, C.-H. Kuan, T.~Li, S.~Zhu, and D.~P. Tsai, ``Broadband
  achromatic optical metasurface devices,'' {\em Nature Communications},
  vol.~8, p.~187, Aug 2017.

\bibitem{Kamali2018}
S.~M. Kamali, E.~Arbabi, A.~Arbabi, and A.~Faraon, ``{A review of dielectric
  optical metasurfaces for wavefront control},'' {\em Nanophotonics}, vol.~7,
  no.~6, pp.~1041--1068, 2018.

\bibitem{Chen2020}
W.~T. Chen, A.~Y. Zhu, and F.~Capasso, ``{Flat optics with
  dispersion-engineered metasurfaces},'' {\em Nat. Rev. Mater.}, vol.~5, no.~8,
  pp.~604--620, 2020.

\bibitem{Yu2011}
N.~Yu, P.~Genevet, M.~A. Kats, F.~Aieta, J.-P. Tetienne, F.~Capasso, and
  Z.~Gaburro, ``Light propagation with phase reflection and refraction,'' {\em
  Science}, vol.~334, pp.~333--337, 2011.

\bibitem{Decker:15}
M.~Decker, I.~Staude, M.~Falkner, J.~Dominguez, D.~N. Neshev, I.~Brener,
  T.~Pertsch, and Y.~S. Kivshar, ``High-efficiency dielectric {H}uygens’
  surfaces,'' {\em Advanced Optical Materials}, vol.~3, pp.~813--820, Feb 2015.

\bibitem{Arbabi2015}
A.~Arbabi, Y.~Horie, M.~Bagheri, and A.~Faraon, ``{Dielectric metasurfaces for
  complete control of phase and polarization with subwavelength spatial
  resolution and high transmission},'' {\em Nat. Nanotechnol.}, vol.~10,
  no.~11, pp.~937--943, 2015.

\bibitem{Lalanne2017}
P.~Lalanne and P.~Chavel, ``{Metalenses at visible wavelengths: past, present,
  perspectives},'' {\em Laser Photonics Rev.}, vol.~11, p.~1600295, May 2017.

\bibitem{Pors2013}
A.~Pors, O.~Albrektsen, I.~P. Radko, and S.~I. Bozhevolnyi, ``{Gap
  plasmon-based metasurfaces for total control of reflected light},'' {\em Sci.
  Rep.}, vol.~3, no.~2155, p.~2155, 2013.

\bibitem{Aieta:12b}
F.~Aieta, P.~Genevet, M.~A. Kats, N.~Yu, R.~Blanchard, Z.~Gaburro, and
  F.~Capasso, ``Aberration-free ultrathin flat lenses and axicons at telecom
  wavelengths based on plasmonic metasurfaces,'' {\em Nano Letters}, vol.~12,
  pp.~4932--4936, Aug 2012.

\bibitem{Arbabi2016}
A.~Arbabi, E.~Arbabi, S.~M. Kamali, Y.~Horie, S.~Han, and A.~Faraon,
  ``{Miniature optical planar camera based on a wide-angle metasurface doublet
  corrected for monochromatic aberrations},'' {\em Nat. Commun.}, vol.~7,
  p.~13682, 2016.

\bibitem{Desiatov2015}
B.~Desiatov, N.~Mazurski, Y.~Fainman, and U.~Levy, ``{Polarization selective
  beam shaping using nanoscale dielectric metasurfaces},'' {\em Opt. Express},
  vol.~23, no.~17, pp.~22611--22618, 2015.

\bibitem{Scheuer:17}
J.~Scheuer, ``Metasurfaces-based holography and beam shaping: engineering the
  phase profile of light,'' {\em Nanophotonics}, vol.~6, pp.~137--152, Jan
  2017.

\bibitem{Khaidarov2020}
E.~Khaidarov, Z.~Liu, R.~Paniagua-Dom{\'{i}}nguez, S.~T. Ha, V.~Valuckas,
  X.~Liang, Y.~Akimov, P.~Bai, C.~E. Png, H.~V. Demir, and A.~I. Kuznetsov,
  ``{Control of LED Emission with Functional Dielectric Metasurfaces},'' {\em
  Laser Photonics Rev.}, vol.~14, no.~1, p.~1900235, 2020.

\bibitem{Mukherjee2023}
S.~Mukherjee, Q.~A.~A. Tanguy, J.~E. Fr{\"{o}}ch, A.~Shanker, K.~F.
  B{\"{o}}hringer, S.~Brunton, and A.~Majumdar, ``{Partially Coherent
  Double-Phase Holography in Visible Wavelength Using Meta-Optics},'' {\em ACS
  Photonics}, vol.~10, pp.~1376--1381, may 2023.

\bibitem{So2023}
S.~So, J.~Mun, J.~Park, and J.~Rho, ``{Revisiting the Design Strategies for
  Metasurfaces: Fundamental Physics, Optimization, and Beyond},'' {\em Adv.
  Mater.}, p.~2206399, apr 2023.

\bibitem{Raza2020}
S.~Raza, ``Slow light using magnetic and electric {M}ie resonances,'' {\em Opt.
  Lett.}, vol.~45, no.~5, pp.~1260--1263, 2020.

\bibitem{Wu:13a}
R.~Wu, P.~Liu, Y.~Zhang, Z.~Zheng, H.~Li, and X.~Liu, ``A mathematical model of
  the single freeform surface design for collimated beam shaping,'' {\em Opt.
  Express}, vol.~21, pp.~20974--20989, Sep 2013.

\bibitem{Aieta:12a}
F.~Aieta, P.~Genevet, N.~Yu, M.~A. Kats, Z.~Gaburro, and F.~Capasso,
  ``Out-of-plane reflection and refraction of light by anisotropic optical
  antenna metasurfaces with phase discontinuities,'' {\em Nano Letters},
  vol.~12, pp.~1702--1706, Feb 2012.

\bibitem{Moreno2020}
I.~Moreno, M.~Avenda{\~{n}}o-Alejo, and C.~P. {Casta\~{n}eda-Almanza},
  ``{Nonimaging metaoptics},'' {\em Opt. Lett.}, vol.~45, no.~10, p.~2744,
  2020.

\bibitem{Castaneda-Almanza:22}
C.~P. {Casta\~{n}eda-Almanza} and I.~Moreno, ``Ray tracing in metasurfaces,''
  {\em Opt. Continuum}, vol.~1, pp.~958--964, May 2022.

\bibitem{Ries:94}
H.~Ries and A.~Rabl, ``Edge-ray principle of nonimaging optics,'' {\em J. Opt.
  Soc. Am. A}, vol.~11, pp.~2627--2632, Oct 1994.

\bibitem{FundamentalsOfPhotonics}
B.~E.~A. Saleh and M.~C. Teich, {\em Beam Optics}, ch.~3, pp.~80--107.
\newblock John Wiley \& Sons, Ltd, second~ed., 2007.

\end{thebibliography}
\newpage

\end{document}